# The Ch-class asteroids: Connecting a visible taxonomic class to a 3-μm band shape

Andrew S. Rivkin[1]*, Cristina A. Thomas[2+], Ellen S. Howell[3*^], Joshua P. Emery[4*]

1. The Johns Hopkins University Applied Physics Laboratory
2. NASA Goddard Space Flight Center,
3. Arecibo Observatory, USRA
4. University of Tennessee

Accepted to The Astronomical Journal

3 November 2015

*Visiting Astronomer at the Infrared Telescope Facility, which is operated by the University of Hawaii under Cooperative Agreement no. NNX-08AE38A with the National Aeronautics and Space Administration, Science Mission Directorate, Planetary Astronomy Program.

+Currently also at the Planetary Science Institute. Work done while supported by an appointment to the NASA Postdoctoral Program, administered by Oak Ridge Associated Universities through a contract with NASA.

^Currently at Lunar and Planetary Laboratory, University of Arizona, Tucson AZ

**Abstract**

Asteroids belonging to the Ch spectral taxonomic class are defined by the presence of an absorption near 0.7 µm, which is interpreted as due to Fe-bearing phyllosilicates. Phyllosilicates also cause strong absorptions in the 3-µm region, as do other hydrated and hydroxylated minerals and $H_2O$ ice. Over the past decade, spectral observations have revealed different 3-µm band shapes the asteroid population. Although a formal taxonomy is yet to be fully established, the "Pallas-type" spectral group is most consistent with the presence of phyllosilicates. If Ch class and Pallas type are both indicative of phyllosilicates, then all Ch-class asteroids should also be Pallas-type. In order to test this hypothesis, we obtained 42 observations of 36 Ch-class asteroids in the 2- to 4-µm spectral region. We found that 88% of the spectra have 3-µm band shapes most consistent with the Pallas-type group. This is the first asteroid class for which such a strong correlation has been found. Because the Ch class is defined by the presence of an absorption near 0.7 µm, this demonstrates that the 0.7-µm band serves not only as a proxy for the presence of a band in the 3-µm region, but specifically for the presence of Pallas-type bands. There is some evidence for a correlation between band depth at 2.95 µm and absolute magnitude and/or albedo. However, we find only weak correlations between 2.95-µm band depth and semi-major axis. The connection between band depths in the 0.7- and 3-µm regions is complex and in need of further investigation.

**Background: Ch asteroids and the 0.7-µm absorption**

In the Bus asteroid taxonomy (Bus and Binzel 2002b), the Cgh and Ch asteroid classes are defined as having a broad absorption feature near 0.7-µm (referred to as the "0.7-µm band" for convenience hereafter), with behavior shortward of 0.55 µm discriminating between the Cgh and Ch classes. The extension of the Bus taxonomy to 2.5 µm by DeMeo et al. (2009) maintained those two classes, with similar definitions. While the shared characteristic of these classes is the 0.7-µm band, the exact band minimum varies from roughly 0.63-0.77 µm in the sample of targets studied by Fornasier et al. (2014). The Cgh class is relatively rare, with Ch asteroids in the SMASS dataset outnumbering them 9:1. For convenience we hereafter refer to members of the Ch and Cgh groups collectively as "Ch asteroids", unless one subset alone is being discussed. There are no significant differences in conclusions considering the groups separately.

The 0.7-µm absorption band seen in the asteroids is also seen in the CM meteorites. The absorption has classically been attributed to $Fe^{2+}$-$Fe^{3+}$ intravalence charge transfer, which is associated with phyllosilicates but not diagnostic of them (Vilas and Gaffey 1989, Vilas 1994, King and Clark 1997). Cloutis et al. (2011) attributes bands near 0.7 µm to saponite group phyllosilicates (band centers 0.59-0.67 µm) and "mixed valence Fe-bearing serpentine group phyllosilicates" (band centers 0.70-0.75 µm). Fraeman et al. (2014) in a study of the spectrum of Phobos and Deimos attributed a band centered near 0.65 µm to phyllosilicates, but noted that a mixture of different-sized µm-scale iron particles in a neutral silicate matrix could also match the 0.65-µm band shape and could not rule it out. Similarly, the ubiquity of phyllosilicates in CM meteorites makes the assignment of the 0.7-µm band straightforward in those meteorites and heavily favors phyllosilicates as the interpretation when the band is seen in asteroids, but we should remain aware that space weathering processes could create mixtures that mimic this feature.

Unlike the Ch asteroids, for which the 0.7-μm band is a defining characteristic, some CM meteorites have no 0.7-μm absorption band. In addition, some non-CM meteorites are found with the band, although such samples discussed in Cloutis et al. (2012) have CM precursors and can be grouped with the CM meteorites for our purposes. This connection provides a partial link between the Ch asteroids and CM meteorites: they are the only asteroids and only meteorites for which absorption bands near 0.7 μm have been identified, as has been noted by Vilas and Gaffey (1989) and Vilas et al. (1994), among others. Carvano et al. (2003) performed least-squares comparisons of the asteroidal spectra in the S3OS2 to laboratory spectra of meteorites and similarly concluded that most asteroids with 0.7-μm bands were "very likely" to be related to the CM2 meteorites. Fornasier et al. (2014) compared the spectra of CM2 meteorites and 80 low albedo asteroids in the 0.4-0.94 μm region, finding the meteorites to have longer wavelength band centers than the asteroids, concluding that the difference might be due to different mineral abundances or the meteorites representing only a subset of hydrated asteroids.

**Background: Pallas-type asteroids and the 3-μm band**

The phyllosilicates that give rise to the absorption bands near 0.7 μm also cause absorptions in the 3-μm region, as do other OH- and/or H2O-bearing minerals. This co-occurrence has led to interest among many workers in the prospect of using the 0.7-μm band as a proxy for the 3-μm band, not least because observations near 0.7 μm are much easier than near 3 μm due to the number of suitable instruments and observing sites and the ease of data collection and reduction (see below).   Vilas (1994) first reported the strong correlation between the 0.7- and 3-µm bands in low-albedo asteroids, and Howell et al. (1999, 2011) have made consistent progress in establishing the usefulness of that correlation.

We have in this century been able to build on the initial reconnaissance phase of 3-µm spectroscopy pioneered by Lebofsky and his co-workers (Lebofsky 1980, Feierberg et al 1985, Lebofsky et al. 1990, Jones et al. 1990, Rivkin et al. 2002).  New instrumentation, analysis techniques, and other factors have allowed us to move beyond a binary presence/absence analysis of the 3-µm band and toward extracting the same kind of information commonly gleaned from 0.4-2.5 µm spectroscopy.

Several studies have established that the carbonaceous chondrite meteorites share particular band shapes, with a strong absorption edge near 2.7 µm due to hydroxyl followed by a relatively linear return to continuum behavior as wavelengths increase.  Earlier laboratory spectral studies (Hiroi et al., 1996, Jones 1989) unavoidably retained some features due to adsorbed telluric water in their spectra, but similar spectral behavior is seen in samples in which telluric water has been removed (Beck et al. 2010, Takir et al. 2013).  However, in those cases where terrestrial adsorbed water is still present, the absorption band is deeper and wider, in the worst cases almost completely obscuring the underlying phyllosilicate absorption.  Studies have shown that the band minimum in the 2.7-2.8 µm region in carbonaceous chondrites is indicative of phyllosilicate composition and degree of aqueous alteration (Sato et al. 1997, Osawa et al. 2005, Beck et al. 2010, Takir et al. 2013).

In contrast to the relatively similar band shapes in the meteorites, the low-albedo asteroid population shows a larger variety of band shapes in the 3-μm region (Rivkin et al. 2002, Rivkin 2010, Takir and Emery 2012, Rivkin et al. 2012, Takir et al. 2015). A formal taxonomy has not yet been created in this wavelength region, but previous workers have identified similar groupings.

Because the Earth's atmosphere is effectively opaque at roughly 2.5-2.85 μm, a band center is not measured from ground-based data. It can, however, be constrained to fall between the wavelengths mentioned, in the atmospheric opaque region. Nevertheless, even without a measurement of the band minimum, many asteroids are seen to have a band shape qualitatively similar to what is seen in the carbonaceous chondrites: a monotonically-increasing, roughly linear reflectance from 2.85 μm (the wavelength at which the sky allows observations) longward until it reaches the continuum value. This shape, first seen in the spectrum of 2 Pallas by Larson et al. (1983), is alternately called "Pallas type" (Rivkin et al. 2012), "sharp" (Takir and Emery 2012), or "checkmark shaped" (Rivkin et al. 2002). In this work we will use the "Pallas type" name, with the understanding that a later formal taxonomy may use a different name. While the lack of data in the area of the band minimum hampers our ability to make precise measurements of phyllosilicate composition, we will discuss several lines of reasoning in the following sections tying the Ch asteroids to the CM meteorites.

**Observations**

The spectra discussed here were collected using the SpeX instrument (Rayner et al. 2003) on the NASA Infrared Telescope Facility (IRTF) through several ongoing projects from 2002-2013. While these projects were not originally intended to constitute a long-term survey, we have retroactively termed the collection the "L-band Main-belt/Near-Earth object Observing Program" (LMNOP) for convenience. Several other results from this program have already been published for individual objects like Vesta (Rivkin et al. 2006a), Ceres (Rivkin et al. 2006b, Rivkin and Volquardsen 2010), Lutetia (Rivkin et al. 2011), and 1996 FG3 (Rivkin et al. 2013). Papers covering other subsets of the data and the survey as a whole are in preparation. Here we concentrate on those objects classified as Ch or Cgh by the SMASS or S3OS2 visible-near IR surveys. The objects in this work were classified as Ch or Cgh by either the Small Mainbelt Asteroid Spectroscopic Survey (SMASS: Bus and Binzel 2002a) or the Small Solar System Objects Spectroscopy Survey (S3OS2: Lazzaro et al. 2004) or both. In the dataset analyzed here, only 3 of the 36 asteroids are Cgh, with the other 33 classified as Ch asteroids.

We used the long-wavelength cross-dispersed mode (LXD), covering 2.0-4.2 μm simultaneously, with a 0.8 arcsecond slit. Integration times were typically 15-20 seconds per exposure, sometimes with 2-3 coadds before a beam switch of 15 arcseconds. Early in the survey, longer exposure times were occasionally used, at the expense of non-linearities at the longest wavelengths. In general, the exposure time of an individual image was set by the brightness of the sky near 4 μm, and the time between beam switches was kept to 2 minutes or shorter in order to allow good sky subtraction under typical conditions. Observing circumstances for each asteroid are provided in Table 1, with details of each night in Table 2.

Data reduction consisted of two main stages. Spextool, a set of IDL-based programs provided by the IRTF, was used to flat-field, wavelength-calibrate, extract, and combine spectra (Cushing et al. 2003). Each asteroid spectrum is divided by each standard star spectrum in the second, post-Spextool phase of reduction, using code provided by E. Volquardsen of the IRTF. This phase corrects for subpixel shifts between objects and uses ATRAN (Lord 1992) to model the atmospheric contribution to each spectrum, using the known observing geometry and fitting the atmospheric precipitable water. The atmospheric precipitable water fits (provided in Table 2) are excellent matches to the values derived from near-simultaneous optical depth measurements at the Caltech Submillimeter Observatory[1]. For each asteroid, every division by a standard star (with atmospheric water removed for asteroid and star) is combined into a final spectrum via weighted average.

Asteroids are warm enough that they have detectable thermal flux in the LXD spectral region. In order to study a reflectance-only spectrum comparable to laboratory spectra of meteorites, this thermal flux must be removed. We used a modified version of the "Standard Thermal Model" (STM) of Lebofsky and Spencer (1989), for which most inputs are known (albedo, solar/Earth distance at time of observation, slope parameter) or are generally agreed upon (emissivity). The one important parameter that is less well known is the beaming parameter ($\eta$), which is a factor that includes surface roughness, shape, thermal inertia, and other effects that are difficult to explicitly model without significantly increasing the complexity of the model. The parameter $\eta$ is equal to 1 for a smooth sphere with zero thermal inertia, and tends to be ~0.8 for low-phase-angle observations of main-belt asteroids, while objects with higher thermal inertias seen at higher phase angles can have η ~1.5 or even higher. We generate several thermal corrections by varying η over its expected range, choosing the model that results in the continuum reflectance near 3.8-4.0 μm that would be extrapolated from the spectral slope near 2.2-2.4 μm and/or a flat continuum beyond ~3.6 μm. If no such fit is found, additional models are generated for a wider range of η values. For the Ch asteroids, which have very little spectral slope, the continuum remains close to the reflectance at 2.2-2.4 μm. The thermal emission amounts to only ~15-20% of the total flux at 3.3 μm and less than 5% near 3.0 μm for low-albedo objects at 2.45 AU, so the thermal correction has a minimal influence on the band depths reported.

---

[1] http://www.cso.caltech.edu/tau/

Finally, the thermally-corrected spectra are binned to a final resolution ($\lambda/\Delta\lambda$) of ~250, representing the pixel width of the slit. In cases where the uncertainties are particularly high, or alternately for display purposes, further binning is done to a final resolution of 60. The figures below will use either of these two types of binned spectra. Where reflectance ratios are quoted, they are computed as the uncertainty-weighted average across the central wavelength +/- 0.01 μm.

Four asteroids were observed and had high-S/N spectra recorded, but are not included in all analyses here because of poor observing conditions, with precipitable water greater than 6 mm, too high to be fully corrected by the reduction software. While comparison to other objects suggests this only has a modest effect on the final analysis, we nevertheless conservatively remove these objects from some discussion.

**Results**

*Thermal properties of the Ch asteroid ensemble:* While the physical meaning of the thermal beaming parameter η is not straightforward, it has been seen to increase with phase angle. Delbo et al. (2007) used thermophysical models to determine the value of η expected from asteroids of various thermal inertias over a wide range of phase angles in order to estimate the thermal inertias of near-Earth asteroids.   Comparing the range of beaming parameters for the asteroid observations in our sample (included in Table 1) to Figure 3 of Delbo et al. shows the Ch sample to be most consistent with the lowest thermal inertias in that work, 15 in SI units. Because the asteroids in our sample are all relatively similar in size (diameters ranging from 61-227 km: Masiero et al. 2011, Usui et al. 2011), this low thermal inertia is not surprising and is consistent with other thermal inertia measurements of large asteroids.  It is also consistent with the thermal inertia of the lunar regolith when the difference in temperature between the Moon and asteroids is taken into account.

*Band shapes in the 3-µm region:* Figure 1 shows 42 spectra obtained of the 36 Ch/Cgh asteroids in the LMNOP. It is evident by inspection that the vast majority of the spectra have 3-µm band shapes similar to that of Pallas in the work of Takir et al. and Rivkin et al. and would be considered "Pallas types" by that standard. There are some cases where data quality is not high, but even in most of these cases, the overall shape can be seen to be more Pallas-like than the other groups.

Figure 2 shows 51 Nemausa, 70 Panopaea, and 48 Doris, three of the highest S/N spectra we obtained. Slight differences in band shape can be seen among them, with a pronounced shoulder from 2.9-3.2 µm in the Panopaea spectrum, a more subtle shoulder in the Nemausa spectrum despite a greater overall band depth, and no obvious shoulder in the Doris spectrum. The spectral modeling required to determine the cause of this shoulder is beyond the scope of this paper.

We use a simple metric, the ratio of reflectance at 3.1 μm to the reflectance at 2.9 μm ($R_{3.1}/R_{2.9}$), to measure the band shape in our sample in a more qualitative manner. This value will be greater than 1 in Pallas-type objects, and less than 1 for the Themis- and Ceres-type objects, which have minima near 3.1 μm. While this metric may be improved upon by later workers, and is likely only appropriate for low albedo objects with very shallow continuum slopes, it provides a way to quickly gauge band shapes for the set of Ch asteroids.

Looking at the set of 42 spectra, 29 have values of $R_{3.1}/R_{2.9} > 1$ σ larger than 1. An additional 8 have values greater than, but within 1 σ of 1. These 37 spectra represent 88% of the sample. Five spectra have $R_{3.1}/R_{2.9}$ values less than 1, and all of these spectra tend to have very large error bars. Four of the five (159 Aemilia, 38 Leda [from 2007], 791 Ani, and 127 Johanna) have $R_{3.1}/R_{2.9}$ values within 0.5 σ of 1, while the fifth (754 Malabar) has a value just over 1 σ less than 1. Malabar is one of the faintest objects presented in this work, and was observed early in the survey, before observing techniques were honed. As a result, we consider its possible deviation from Pallas-type spectra as unproven, and expect higher-quality data to reveal a Pallas-type spectrum. The other objects in the group of 5 also have mitigating circumstances: 791 Ani was classified as Ch in S3OS2, but inspection of the spectrum leaves doubt as to whether or not a band is present at 0.7 μm, and Fornasier et al. (2014) classified that spectrum as anhydrous. Ani might therefore not be properly included in the sample of Ch asteroids. The spectrum of Leda from 2008 is one of the 29 objects with $R_{3.1}/R_{2.9} > 1$ by more than 1 σ, opening the possibility that the 2007 spectrum was not measuring the same portion of the surface that has the 0.7-μm band. The spectrum of 159 Aemilia is of extremely low quality. Finally, the night on which 127 Johanna was observed had very high precipitable water, and inspection of its spectrum in Figure 1 suggests that the 2.9-μm reflectance point may have been particularly affected—the rest of the spectrum appears consistent with a Pallas-type spectrum.

Whereas the sample of Ch/Cgh asteroids have Pallas-type spectra (or have reasonable explanations related to data quality for not having them), and thus asteroids with a 0.7-μm band may also be expected to have Pallas-type 3-μm bands, we caution that other combinations are not necessarily true: Not every object with a Pallas-type 3-μm band has a 0.7-μm band, and not every CM meteorite has a 0.7-μm band. Similarly, although the Ch asteroids with high S/N spectra taken on good observing nights all have Pallas-type bands, other asteroid classes have a mixture of 3-μm band shapes, including Pallas-type ones. We also note that the Ch/Cgh population is likely continuous with other subsets of the C-complex population, and the boundary between Ch and C asteroids is fuzzy, if only due to the imperfect nature of real data and the possibility that below some threshold, 0.7-μm bands may not be identified due to a too-shallow band depth (and that that threshold may vary with observing conditions). Nevertheless, the trends of band depth and reflectance ratios with orbital and physical properties provide useful information and constraints on the Ch population, discussed here.

**Discussion**

*Hydrogen:Silicon ratio:* Rivkin et al. (2003), following work by Miyamoto and Zolensky (1994) and Sato et al (1997), established a means of estimating the H/Si ratio of asteroids using their 2.9 µm/2.5 µm and 3.2 µm/2.5 µm reflectance ratios.   More recent spectral work by Beck et al. and Takir et al. demonstrated that terrestrial water has a significant effect on the 3-µm region in laboratory spectra of carbonaceous chondrites. While those workers heated samples to remove terrestrial adsorbed water, the degree to which adsorbed water affected the earlier laboratory measurements is not clear. On the other hand, it seems likely that the calculation of the H:Si ratio does not need to be rederived using new laboratory data since the Rivkin et al. work used both chemically bound and free water in their calculations (reported as "H2O+" and "H2O-" in Jarosewich 1990), not just bound water.

    Figure 3 shows the H:Si ratio for the set of Ch asteroids in the LMNOP with high-quality spectra from good-weather (PW < 6 mm) nights, and also for the meteorite spectra published by Takir et al (stippled squares) and the CM meteorites reported in Jarosewich (1990, open squares).   The values for each group are sorted by H:Si value for easier viewing in the figure.  The H:Si values for the Jarosewich data set are calculated from the listed $SiO_2$ and $H_2O+$ (bound water) values, whereas they are calculated from the spectra of the Takir et al data set. The two groups of meteorites span similar ranges in H:Si, including some surprisingly low values. The Ch asteroids cover a similar but smaller range compared to the meteorites. The mean H:Si ratio for all high-quality Ch asteroid observations is 1.56 +/- 0.76. Excluding those taken on poor-quality nights results in a mean H:Si ratio of 1.56 +/- 0.79, indicating the weather is not significantly affecting results.   For comparison, performing the same calculation on the set of 14 S-complex asteroids observed in the LMNOP (Rivkin et al. in prep) gives a mean H:Si ratio of 0.05 +/- 0.33, consistent with our expectations that the S asteroids are basically anhydrous.  The consistency of the H:Si ratios

calculated for the Takir et al. and Jarosewich data despite the different means of generating the elemental ratios supports our contention that the equations relating spectral properties with H:Si ratio do not need to be rederived. The consistency of the asteroids with the meteorites further strengthens the proposed link between the Ch asteroids and CM meteorites.

We note that while it may be tempting to work from first principles to compute an expected H:Si ratio or convert the H:Si ratio into a phyllosilicate fraction, this is not necessarily straightforward. Phyllosilicate groups like serpentine and chlorite have stoichiometric H:Si ratios of ~2-2.5, but saponite group minerals on Earth can have variable amounts of water of crystallization, so they can have H:Si ratios ranging from 0.67 to an arbitrarily high value. In addition, minerals like tochilinite have hydrogen but no silicon. Therefore, plausible H:Si ratios for CM meteorites can cover a relatively wide range, as we see here. For completeness, however, we note that Howard et al. (2013) found that CM2 meteorites had a relatively narrow range of phyllosilicate abundance of 73-79%, with an average of 75%, and similarly narrow ranges in olivine and pyroxene abundances. If all the hydrogen in CM2 meteorites is considered to be in phyllosilicates, and all the silicon in phyllosilicate, olivine, and pyroxene, the resulting H:Si ratio expected is ~1.8-1.9. That value is higher than, but within 1-$\sigma$ of, the mean value of the Ch asteroids, and ten asteroids within the sample have H:Si ratios of 1.8 or higher. Looking at the scatter within both the Ch asteroid and the CM meteorite populations, they appear to be consistent with one another in several ways in terms of H:Si ratio.

*Water vs. hydroxyl in Ch asteroids and CM meteorites*: Milliken and Mustard (2005) investigated various approaches to quantifying water concentrations in hydrated minerals from reflectance spectra. The normalized optical path length (NOPL) technique is well-suited to the asteroid spectra discussed here. Milliken and Mustard were interested in $H_2O$ per se rather than OH, and they explicitly excluded absorptions solely due to hydroxyl in their calculations. Using the relation between NOPL and water concentration suggests there is very little adsorbed water in the Ch asteroids, with even the deepest band depths indicating only ~0.7% water. This is not surprising, since we expect most of the hydrogen detected on the Ch asteroid surfaces to be part of hydroxyl groups in phyllosilicates or other minerals, and loosely bound water is thought unlikely to remain after extended exposure to vacuum. Applying the Milliken and Mustard relation to the heated/dried meteorite spectra of Takir et al. results in concentrations similar to what is seen in the asteroids.

*Trends in Ch population:* Figures 4-11 show the band depth (BD) trends in the Ch population. The band depth at 2.95 µm is used in figures unless otherwise noted. Several of the correlations we investigated are significant at the 95% confidence level: 2.95 µm BD with 3.2 µm BD (discussed in more detail below), 2.95 BD vs. absolute magnitude (H), 2.95 µm BD vs. albedo, 2.95 µm BD vs. diameter, and albedo vs. diameter. We take albedo and diameter values (Table 3) from the WISE dataset (Masiero et al. 2011), augmented by AKARI data (Usui et al. 2011) where WISE values are unavailable. The range of diameters in our sample is 61-227 km, with the median diameter 127 km. The range of albedos is 0.033-0.10, with the median albedo 0.057. The band center for the 0.7-µm absorption ranges from 0.687 to 0.740 μm in the sample of asteroids considered here (Bus and Binzel 2002a, Lazzaro et al. 2004).

Two of the correlations (2.95 BD vs. 0.7 μm band center, and 2.95 μm BD vs. semi-major axis) are not significant at the 95% level. The correlation of 2.95 BD with 0.7 μm BD is also not significant at the 95% level, but is of general interest and is also discussed in more detail below.

The 2.95 μm BD is very poorly correlated with semi-major axis (Figure 4), suggesting the population is well mixed in the asteroid belt. This is consistent with the results of Rivkin (2012) who found the fraction of Ch asteroids within the C complex to be relatively steady with semi-major axis. It is also consistent with Fornasier et al. (2014), who found no correlation between orbital elements and fraction of hydrated asteroids.

The correlations between 2.95 μm BD, albedo, and diameter (as well as H, which is a function of diameter and albedo and thus will not be explicitly discussed) shown in Figures 5-7 are all intertwined. Given the strong correlation between albedo and diameter in this sample of spectra, it is not immediately obvious whether the BD-albedo correlation is an artifact of the BD-diameter correlation or vice versa, or if there is a true correlation between BD and both of these parameters. Looking at the WISE dataset does not necessarily clarify the picture. The 117 Bus-classified Ch asteroids in the WISE dataset (Masiero et al. 2011) taken as a whole do not show any correlation at all between albedo and diameter, suggesting that the subset of asteroids studied here is biased in some way. However, when restricting the WISE-observed Ch asteroids to the 62 objects > 60 km in diameter, similar to the size range of the asteroids in our sample (Table 3), the correlation between albedo and diameter jumps dramatically, and is significant at the 99% level.

Because this change in albedo vs. diameter correlation strength occurs entirely within the WISE data, it is beyond the scope of this work to investigate in depth, but it is puzzling. When removing the albedo-diameter trend and correcting for it, the BD-diameter correlation coefficient decreases and the significance drops below the 90% level, while the BD-albedo significance only drops a little below 95%, suggesting the latter correlation is more significant.

Both a 2.95 µm BD vs. diameter correlation and a 2.95 µm BD vs. albedo correlation are consistent with trends found by others. Clark (1983) found that both the 2.8-µm BD and the reflectance of clay-opaque mixtures varied inversely with log(weight fraction of opaques) in the reflectance range relevant to Ch asteroids. As a result, reflectance and 2.8-µm BD should have a linear relationship with each other. Because the SpeX data do not cover the 2.8-µm region, we cannot make a direct comparison, however we expect the 2.8-µm and 2.95-µm BD to be closely related. Relatedly, Vilas (1994) found a correlation in low-albedo Tholen spectral classes between the fraction of a class showing a 0.7-µm band and mean albedo of that class, arguing that the aqueous alteration process would increase albedos.

A correlation between band depth and diameter has been considered by several workers, but little consensus has been reached. Jones et al. (1990) found larger bodies to have deeper 3-µm bands, but his sample spanned several low-albedo classes and Ceres and Pallas dominated the trend. Takir and Emery (2012) found a correlation between 2.9-µm band depth and diameter for a sample of 13 targets (including 7 Ch asteroids) between 100-300 km diameter, attributing it to hydrothermal circulation causing aqueous alteration in the regolith. While we use 2.95-µm band depth in this work, we find a similar but weaker correlation than that found by Takir and Emery if we use 2.9-µm band depth.

Vilas and Sykes (1996) found that the fraction of asteroids with 0.7-µm bands decreased in smaller size bins (a trend confirmed by Rivkin 2012 to smaller sizes and also seen by Fornasier et al. 2014), which they attributed to smaller objects being more likely to represent samples of heated/dehydrated interiors, while larger objects were more likely to retain original surfaces that never reached high temperatures. Conversely, Carvano et al. (2003) found no correlation between their RRE parameter (which measures 0.7-µm band depth) and diameter among objects in the S3OS2 dataset, which might be expected if smaller objects are more likely to represent interiors and larger objects more likely to retain original surfaces. Construction of scenarios to reconcile lack of a correlation between 0.7-µm band depth and diameter with the existence of a hydrated fraction-diameter correlation are beyond the scope of this paper.

*Spectral correlations*: Two correlations between band depths were investigated, that between the 2.95 µm BD and 3.2 µm BD (Figure 8), and between the 2.95 µm BD and 0.7 µm BD (Figures 9-11). Given that same band encompasses both 2.95 µm and 3.2 µm, it is not surprising that they have a very close correlation, significant at the 99.9%+ level. Still, it does strengthen the conclusion that all of these asteroids, selected based on visible-near IR spectra, also have similar 3-µm band shapes.

The correlation between the 0.7- and 2.95-µm BD is more difficult to interpret overall, as will be detailed in the next section. The asteroid sample when considered alone shows a correlation coefficient very close to zero between these two band depths.

*Correlations with meteorites:* The correlations investigated above can be extended by including the CM meteorite population in the sample. The 2.95 μm BD-3.2 μm BD plot (Figure 8) shows the CM meteorites falling along the same trend as the Ch asteroids. The meteorites are divided in two groups—those drawn from the RELAB database, and those from Takir et al. The latter group has had terrestrial water removed and so falls closer to the origin. As seen in the H:Si discussion, the Takir et al. meteorites fall among the Ch asteroids on this trend.

Interestingly, however, while the BD plots in the 3-μm region appear as expected, the 0.7 μm BD-2.95 μm BD plot does not. Instead, the group of asteroids and meteorites show a single strong correlation with > 99.9% confidence if taken together (Figure 10). Those meteorites for which both the 0.7-μm band and the 3-μm band are present and for which measurements are available appear to continue the trend found in the Ch asteroids. While at first blush this might be expected, we note again that these meteorites have by and large not had terrestrial water removed from their spectra and thus likely have deeper 3-μm bands than we might expect if they were in space. The work of Milliken and Mustard and Takir et al. suggests that band depths in the 3-μm region could decrease by a factor of 50% or more if terrestrial water is removed. However, we would not expect the 0.7-μm band, caused by a charge transfer in iron rather than directly by water or hydroxyl, to be affected by mere removal of adsorbed water. Therefore, we might expect the meteorites measured in ambient conditions (representing 26 of the 30 meteorite measurements) to move horizontally to the left on Fig 10, off of the trend created by the asteroids. Figure 11 shows the combined set of asteroid and meteorite measurements under different conditions. The "ambient" set includes the asteroids, Takir et al. measurements, and RELAB spectra as found in the database. The "dry" set reduces the 2.95-μm BD for the RELAB spectra to 45% of the ambient value as an estimate of what their band depths may be like if measured under conditions similar to the Takir et al. data. The apparent correlation between 0.7-μm and 2.95-μm BD is significantly reduced after this estimated correction is made, and the significance drops below 99%. However, the statistical significance remains above 90%. It is not yet clear what the relationship between the 0.7-μm and 2.95-μm band depths is in detail, underscoring the need for laboratory measurements to be taken under dry conditions.

Because the correction presented in Figure 11 is somewhat ad hoc, we stress that without simultaneous measurements of both of these spectral regions in meteorites, the exact nature of a combined asteroid/meteorite trend is still somewhat speculative.

*A feature near 2.33 μm?:* The spectra in Figure 2 also show evidence of an absorption near 2.33 μm with a band depth of ~1%. Figure 12 focuses on the 2.2-2.5 μm region and includes six targets from our sample, with some targets showing obvious absorptions and others less-conclusive evidence. When present, this feature is consistent with those attributed to mixed-valence Fe-bearing serpentines or saponite group phyllosilicates in CM chondrites by Cloutis et al. (2011), and consistent in band depths and positions with the "2.3-μm band" seen by Takir et al. (2013) in their sample of CM and CI carbonaceous chondrites. This provides an additional link between the Ch asteroids and CM meteorites, although it is not by itself as secure as the presence of the 0.7-μm band or the 3-μm band shape.

**Conclusions**

The 2-4 µm spectra of 36 asteroids in the Ch/Cgh taxonomic classes were obtained using the SpeX instrument on the IRTF in order to study their hydrated mineralogy. Thermal models produced as part of data reduction suggest that as a group the Ch/Cgh asteroids in this sample have thermal inertias consistent with other large asteroids of other compositions.

While they were selected on the basis of a spectral class defined from 0.5-1.0 µm data, we find the Ch/Cgh asteroids share very similar band shapes in the 3-µm region, called "Pallas-type", "sharp", or "checkmark-shaped" in the literature. The presence of the 0.7-µm absorption band on the Ch/Cgh asteroids has been used as a link between them and the CM meteorites. Estimates of their H:Si ratios and measures of spectral ratios from this work show them to be similar to the CM meteorites in several additional ways, strengthening the proposed link. The presence of a 0.7-µm band is evidence not only that the Ch/Cgh asteroids have hydrated minerals, but that each individual asteroid for which 3-µm band shapes can be determined has the kind of hydrated minerals seen in the CM meteorites.

Correlations in the Ch/Cgh population are present but can be difficult to interpret. Larger objects in the dataset appear to have deeper band depths in the 3-µm region, as do higher-albedo ones. There is little evidence of a connection between band depth and semi-major axis. The correlation between band depths in the 0.7- and 3-µm region in the asteroids rests on the inclusion of a single object. Conversely, the meteorites and asteroids taken as one group has a very robust correlation between those band depths, though this correlation may paradoxically depend critically upon the meteorites being measured in ambient atmospheric conditions rather than dry, vacuum, asteroid-like ones, and therefore invalid in the conditions that are expected on real objects.

*Acknowledgements:* ASR gratefully acknowledges consistent support from the NASA Planetary Astronomy program, including grants NAG5-13221, NNG05GR60G, NNX09AB45G, and NNX14AJ39G.  ESH was partially supported by NSF AST-1109855 and NASA NEOO NNX12AF24G.  CAT was supported by an appointment to the NASA Postdoctoral Program at Goddard Space Flight Center, administered by Oak Ridge Associated Universities through a contract with NASA. This material is based upon work at the Caltech Submillimeter Observatory, which was operated by the California Institute of Technology under cooperative agreement with the National Science Foundation (AST-0838261). We acknowledge the sacred nature of Mauna Kea to many Hawaiians, and our status as guests who have been privileged to work there.  Many thanks to the stalwart telescope operators of the IRTF who were instrumental in taking these data through the years, and to Bobby Bus and Eric Volquardsen for developing the "ATRAN part" of the data reduction. ASR thanks Ashwin Raghavachari for his supporting work on the database. This research utilizes spectra acquired by Takahiro Hiroi with the NASA RELAB facility at Brown University. Thanks to an anonymous reviewer who we hope is charitably disposed toward the published paper!

| Date | Asteroid | V Mag | Int Time (s) | Phase Angle | η |
|---|---|---|---|---|---|
| 13-Sep-07 | 13 Egeria | 10.86 | 1080 | 13.2 | 0.8 |
| 21-Feb-07 | 19 Fortuna | 11.39 | 1000 | 11.4 | 1.05 |
| 8-Jan-13 | 34 Circe | 11.96 | 720 | 9.2 | 0.9 |
| 11-Jun-07 | 38 Leda | 12.65 | 720 | 2.6 | 0.9 |
| 21-Jun-08 | 38 Leda | 13.67 | 960 | 17.6 | 0.83 |
| 29-Aug-04 | 41 Daphne | 11.08 | 1200 | 10.4 | 0.85 |
| 15 May 2008** | 41 Daphne | 10 | 1440 | 19.5 | 0.75 |
| 2-Mar-03 | 48 Doris | 11.56 | 720 | 6 | 0.95 |
| 22-Feb-07 | 51 Nemausa | 10.14 | 800 | 6.5 | 0.95 |
| 28-Feb-03 | 66 Maja | 13.69 | 1600 | 8.9 | 0.9 |
| 30-Oct-10 | 70 Panopaea | 11.38 | 1680 | 3.3 | 0.85 |
| 30-Apr-07 | 78 Diana | 12.47 | 1440 | 28.1 | 0.85 |
| 15-May-06 | 105 Artemis | 11.77 | 1440 | 23.9 | 0.88 |
| 30-Jan-12 | 109 Felicitas | 11.88 | 1440 | 20 | 0.85 |
| 23 Jul 2013** | 127 Johanna | 12.7 | 1260 | 8.2 | 0.95 |
| 14-May-06 | 130 Elektra | 13.12 | 2160 | 11.5 | 0.76 |
| 12-Jun-07 | 130 Elektra | 12.38 | 1200 | 8.8 | 0.95 |
| 25-Aug-13 | 130 Elektra | 12.01 | 1740 | 15.5 | 0.93 |
| 30-Oct-10 | 144 Vibilia | 10.38 | 1440 | 9 | 0.85 |
| 21-Jun-04 | 156 Xanthippe | 11.81 | 720 | 17.8 | 0.9 |
| 21-Jun-04 | 159 Aemilia | 13.22 | 1200 | 7.8 | 1.4 |
| 27-Feb-03 | 163 Erigone | 11.98 | 1040 | 10.2 | 1 |
| 27-Jun-12 | 168 Sibylla | 13.22 | 1620 | 12.5 | 0.87 |
| 22-Nov-07 | 176 Iduna | 13.03 | 2400 | 21.2 | 0.95 |
| 17-Sep-06 | 200 Dynamene | 11.88 | 1200 | 7.9 | 0.87 |

| Date | Target | V mag | Integration time (s) | Phase angle (°) | η |
|---|---|---|---|---|---|
| 20-Mar-02 | 207 Hedda | 13.02 | 1560 | 13.8 | 0.88 |
| 2-Mar-03 | 211 Isolda | 12.01 | 720 | 2.5 | 0.8 |
| 22-Sep-05 | 345 Tercidina | 12.11 | 1920 | 12.5 | 0.88 |
| 28-Feb-03 | 404 Arsinoe | 11.94 | 1520 | 10.3 | 0.88 |
| 12 Jun 2007** | 405 Thia | 12.12 | 300 | 17.6 | 0.9 |
| 30-Oct-10 | 407 Arachne | 12.38 | 1440 | 9.9 | 0.85 |
| 7-Sep-06 | 554 Peraga | 12.21 | 1080 | 21.1 | 0.88 |
| 17-Sep-06 | 554 Peraga | 11.94 | 1200 | 17.7 | 0.9 |
| 22-Jul-09 | 554 Peraga | 12.37 | 2400 | 4.1 | 0.93 |
| 23 Jul 2013** | 576 Emanuela | 12.63 | 1440 | 5.3 | 0.95 |
| 12-Sep-07 | 602 Marianna | 12.43 | 1320 | 22.7 | 0.85 |
| 20-Mar-02 | 654 Zelinda | 11.52 | 1160 | 32.4 | 0.85 |
| 7 Sep 2009** | 694 Ekard | 11.59 | 1080 | 24.5 | 0.93 |
| 30-Aug-04 | 754 Malabar | 13.69 | 1200 | 4.8 | 0.85 |
| 16-May-06 | 776 Berbericia | 12.36 | 1440 | 3.6 | 1.05 |
| 21-Jun-08 | 791 Ani | 13.25 | 1200 | 13.1 | 0.93 |
| 26-Jun-13 | 1467 Mashona | 13.93 | 720 | 19.7 | 1 |

**Table 1:** Observing circumstances for targets. Integration time is calculated from the number of frames, co-adds per frame, and exposure time per co-add. V magnitude and phase angle were provided by the JPL Horizons ephemeris service. The beaming parameter (η) is the best-fit value as described in the Observations section. **Night had precipitable water values > 6 mm.

| Date | Standard Stars | PW Range (mm) |
|---|---|---|
| 20 March 2002 | SAO 138636, SAO 183975, SAO 161514, SAO 158315, SAO 117108 | 0.9-1.2 |
| 27 February 2003 | HD 119550, SAO 119691, L102-1081 | 1.6-2.0 |
| 28 February 2003 | L102-1081, HD 98562, HD 131715 | 0.8-1.6 |
| 2 March 2003 | SAO 65083, L102-1081 | 0.6-0.9 |
| 21 June 2004 | 16 Cyg B, HD 137781, HD 184700 | 2.6-6* (6.7) |
| 29 August 2004 | HD 184700, HD 174466, HD 156802, HD 213199 | 1.4-1.6 |
| 30 August 2004 | 16 Cyg B, HD 184700, HD 170717, HD 151450 | 1.0-1.7 |
| 22 September 2005 | 16 Cyg B, HD 211476 | 3.1-3.9 |
| 14 May 2006 | HD 128596, HD 602981, HD115762 | 1.4-1.8 |
| 15 May 2006 | HD 107146, HD 119638, HD 60298 | 0.8-1.3 |
| 16 May 2006 | HD 115642, HD 144821, HD 107146 | 0.8-1.1 |
| 7 September 2006 | HD 12846 | 1.4-1.5 |
| 17 September 2006 | HD 12846, HD 177082, HD 193193, HD 211476, HD 224383 | 3.1-3.9 |
| 21 February 2007 | HD 107146, HD 110747 | 2.9-3.0 |
| 22 February 2007 | HD 107146, HD 128596, HD 60298, HD 77730, HD 88371 | 2.2-2.9 |
| 30 April 2007 | HD 102196, HD 119638, HD 77730 | 0.9-1.1 |
| 11 June 2007 | HD 151928, HD 184700, HD 189499, HD 205027 | 0.5-1.4 |
| 12 June 2007 | HD 140990, HD 198273 | 4.1-6* |
| 12 September 2007 | HD 211476, HD 23169, HD 31867, HD 377 | 2.2-3.3 |
| 13 September 2007 | HD 13043, HD 211476, HD 26749, HD 377 | 1.9-2.4 |

| Date | Objects | Precipitable Water (mm) |
|---|---|---|
| 22 November 2007 | HD 205027, HD 377 | 3.5-4.5 |
| 15 May 2008 | HD 110747, HD 114821, HD 153631, HD 88371, | 4.3-6* (6.0) |
| 21 June 2008 | HD 153631, HD 198273 | 1.9-2.8 |
| 22 July 2009 | HD 126868, HR 6697, HR 206827, HD 1835 | 3.5-4.9 |
| 7 September 2009 | HD 205027 | 5.0-5.9 |
| 30 October 2010 | HD 23169, HD 60298, HR 1024, HR 9107 | 1.1-1.2 |
| 30 January 2012 | HD 12846, HD 60298 | 1.1-1.4 |
| 27 June 2012 | 16 Cyg B, HD 128596, HD 203311 | 0.7-0.8 |
| 8 January 2013 | HD 42160, HD 77730, Hya 64 | 1.0-1.8 |
| 26 June 2013 | HD 184700, HD 223238 | 1.0-2.3 |
| 23 July 2013 | HD 177911, HD 193193, HD 223238 | 6* (8.2) |
| 25 August 2013 | SA155-271, SAO 142780, SAO 161608 | 0.6-1.0 |

**Table 2:** Observing nights. Precipitable water is range calculated for all objects during the night. Entries with 6* indicate precipitable water amount went higher than 6 mm during night. Numbers in parentheses indicate highest precipitable water on those nights as calculated using CSO 225 GHz measurements (see footnote 1 for reference).

| Asteroid | Class | Diameter (km) | Albedo | Semi-major axis (AU) |
|---|---|---|---|---|
| 13 Egeria | Ch | 227 | 0.069 | 2.57 |
| 19 Fortuna | Ch | 223 | 0.050 | 2.44 |
| 34 Circe | Ch | 113.2 | 0.054 | 2.69 |
| 38 Leda | Cgh | 116 | 0.062 | 2.74 |
| 41 Daphne | Ch | *179.6* | *0.078* | 2.76 |
| 48 Doris | Ch | 223.4 | 0.062 | 3.11 |
| 51 Nemausa | Ch | 142.6 | 0.10 | 2.37 |
| 66 Maja | Ch | *71.8* | *0.062* | 2.65 |
| 70 Panopaea | Ch | 139 | 0.040 | 2.61 |
| 78 Diana | Ch | *126.5* | *0.064* | 2.62 |
| 105 Artemis | Ch | 119 | 0.047 | 2.37 |
| 109 Felicitas | Ch | 89 | 0.071 | 2.70 |
| 127 Johanna | Ch | *114.2* | *0.065* | 2.76 |
| 130 Elektra | Ch | 198.9 | 0.071 | 3.12 |
| 144 Vibilia | Ch | 142.2 | 0.06 | 2.65 |
| 156 Xanthippe | Ch | 110.7 | 0.050 | 2.73 |
| 159 Aemilia | Ch | 127.4 | 0.061 | 3.10 |
| 163 Erigone | Ch | 81.6 | 0.033 | 2.37 |
| 168 Sibylla | Ch | 144 | 0.057 | 3.38 |
| 176 Iduna | Ch | 122.1 | 0.082 | 3.19 |
| 200 Dynamene | Ch | 122.1 | 0.082 | 2.74 |
| 207 Hedda | Ch | 130.5 | 0.052 | 2.28 |
| 211 Isolda | Ch | 143 | 0.060 | 3.04 |
| 345 Tercidina | Ch | 99 | 0.059 | 2.33 |
| 404 Arsinoe | Ch | 98.7 | 0.045 | 2.59 |
| 405 Thia | Ch | 125 | 0.047 | 2.58 |
| 407 Arachne | Ch* | *97.54* | *0.052* | 2.62 |
| 554 Peraga | Ch | *96.98* | *0.049* | 2.37 |
| 576 Emanuela | Cgh* | 77.2 | 0.052 | 2.99 |
| 602 Marianna | Ch* | 126.8 | 0.052 | 3.09 |
| 654 Zelinda | Ch | 127 | 0.043 | 2.30 |
| 694 Ekard | Ch* | *92.11* | *0.045* | 2.67 |
| 754 Malabar | Ch | *102.8* | *0.043* | 2.99 |
| 776 Berbericia | Cgh | 151.1 | 0.066 | 2.93 |
| 791 Ani | Ch* | 82.5 | 0.052 | 3.12 |
| 1467 Mashona | Ch* | 104.1 | 0.061 | 3.38 |

**Table 3:** Physical properties of target asteroids. *Italicized* values are taken from AKARI survey (Usui et al. 2011), others from WISE survey (Masiero et al. 2011). Asteroid classes are taken from SMASS survey (Bus and Binzel 2002b), save those with *, which are taken from the S3OS2 survey (Lazzaro et al. 2004).

| Meteorite | Source | Wavelengths | File | Reference |
|---|---|---|---|---|
| ALH84033,21 | RELAB | 0.3-25 μm | ncmp14 | Hiroi et al. (1996b) |
| Y74642 | RELAB | 0.3-25 μm | ncmb75 | Hiroi et al. (1996a) |
| Bells | RELAB | 0.3-25 μm | ncmb53 | Hiroi et al. (1996a) |
| Y-82098 | RELAB | 0.3-25 μm | ncmp08 | Hiroi et al. (1997) |
| EET87522,29 | RELAB | 0.3-25 μm | ncmp21 | Hiroi et al. (1997) |
| Y-793321 | RELAB | 0.3-25 μm | ncmp13 | Hiroi et al. (1997) |
| MAC88100,30 | RELAB | 0.3-25 μm | ncmp22 | Hiroi et al. (1997) |
| Y-791198 | RELAB | 0.3-25 μm | ncmp12 | Hiroi et al. (1996b) |
| Y-86789 | RELAB | 0.3-25 μm | ncmp10 | Hiroi et al. (1996b) |
| Y74662 | RELAB | 0.3-25 μm | ncmb76 | Hiroi et al. (1996a) |
| Y-74662,101 | RELAB | 0.3-25 μm | ncmp11 | Hiroi et al. (1996a) |
| Murray | RELAB | 0.3-25 μm | ncmb56 | Hiroi et al. (1996a) |
| ALHA81002 | RELAB | 0.3-25 μm | ncmb50 | Hiroi et al. (1996a) |
| Mighei | RELAB | 0.3-25 μm | ncmb55 ,bkr1ma62, bkr1ma63 | Hiroi et al. (1996a) |
| ALHA77306,45 | RELAB | 0.3-25 μm | ncmp19 | Hiroi et al. (1996b) |
| Nogoya | RELAB | 0.3-25 μm | ncmb62 | Hiroi et al. (1996a) |
| ALH85013,32 | RELAB | 0.3-25 μm | ncmp23 | Hiroi et al. (1997) |
| Murchison | RELAB | 0.3-25 μm | ncmb64 | Hiroi et al. (1996a) |
| LEW87022,26 | RELAB | 0.3-25 μm | ncmp20 | Hiroi et al. (1996b) |
| LEW90500 | RELAB | 0.3-25 μm | ncmb54 | Hiroi et al. (1996a) |
| GRO85202,16 | RELAB | 0.3-25 μm | ncmp17 | Hiroi et al. (1996b) |
| LEW87148,15 | RELAB | 0.3-25 μm | ncmp16 | Hiroi et al. (1996b) |
| ALH84029 | RELAB | 0.3-25 μm | ncmb52 | Hiroi et al. (1996a) |
| LEW85311,36 | RELAB | 0.3-25 μm | ncmp24 | Hiroi et al. (1997) |
| ALH83100 | RELAB | 0.3-25 μm | ncmb51 | Hiroi et al. (1996a) |
| ALH84044,16 | RELAB | 0.3-25 μm | ncmp15 | Hiroi et al. (1996b) |
| Bells | Takir | 0.4-4 μm | | Takir et al. (2013) |
| Ivuna | Takir | 0.4-4 μm | | Takir et al. (2013) |
| LAP 022777 | Takir | 0.4-4 μm | | Takir et al. (2013) |
| MET 00639400 | Takir | 0.4-4 μm | | Takir et al. (2013) |
| MAC 02606 | Takir | 0.4-4 μm | | Takir et al. (2013) |
| MIL 077000 | Takir | 0.4-4 μm | | Takir et al. (2013) |
| QUE 97990 | Takir | 0.4-4 μm | | Takir et al. (2013) |
| QUE 99038 | Takir | 0.4-4 μm | | Takir et al. (2013) |
| Cold Bokkeveld | Takir | 0.4-4 μm | | Takir et al. (2013) |
| LAP 03786 | Takir | 0.4-4 μm | | Takir et al. (2013) |
| Dhofar 225 | RELAB | 0.3-25 μm | bkr1ma78 | Hiroi, PI |
| MET 00639 | RELAB | 0.3-25 μm | bkr1ph032 | Hiroi, PI |
| WIS 91600 | RELAB | 0.3-25 μm | bkr1ph033 | Hiroi et al. (2005) |
| QUE97077 | RELAB | 0.3-50 μm | bkr1ph052 | Hiroi, PI |

**Table 4:** Meteorites with spectra used in this study. "Hiroi, PI" means that these were present in the RELAB public database and were obtained by Takahiro Hiroi but have not been featured in a publication.

**Figure Captions**

**Figure 1:** Spectra of Ch asteroids. The asteroid number is included in each box, with a letter appended if an object was observed more than once, in order of observation. See Table 1 for details of observations. Error bars are 1-σ.

**Figure 2:** Three Ch asteroids, offset from one another for clarity, show examples of a shoulder near 3.1 µm of varying depth. This shoulder is likely due to slight mineralogical differences between the asteroids, though detailed modeling would be necessary to determine the exact cause. Error bars are 1-σ.

**Figure 3:** H:Si ratios calculated from high-S/N spectra of Ch asteroids on nights with low precipitable water, compared to calculated H:Si ratios from spectra of CM meteorites measured by Takir et al. (2013) and reported H:Si measurements of CM meteorites from Jarosewich (1990). All cover a similar range of values. Error bars are 1-σ.

**Figure 4:** 2.95-µm Band Depth vs. Semi-Major Axis. There is no statistically significant trend in band depth of Ch asteroids with distance from the Sun, and the greatest variation is seen in the outer asteroid belt. Error bars are 1-σ.

**Figure 5:** Correlation between albedo and 2.95-µm band depth for the sample of Ch asteroids studied in this work. There is a tendency for higher-albedo objects to have deeper bands, significant at the 95% level. Error bars are 1-σ.

**Figure 6:** Similar to Figure 9, but considering band depth and diameter. There is a relatively weak tendency for larger objects to have deeper band depths, though the greatest variation in band depth in the sample is found from ~70-120 km. Error bars are 1-σ.

**Figure 7:** Correlation between diameter and albedo for the sample studied here. The albedos and diameters are drawn from the WISE and AKARI surveys and are not generated by this work. The correlation seen is stronger than seen for either of these parameters and 2.95-µm band depth. While curious, the cause of this correlation resides outside the scope of this work.

**Figure 8:** The 2.95-μm band depth vs. 3.2-μm band depth trend in CM meteorites measured in ambient conditions (diamonds) is consistent with those measured by Takir et al. in vacuum after heating to drive off terrestrial water (triangles). The trend in Ch asteroids (squares) is consistent with the CM trends, though with greater scatter. The scatter may be due in part to real spectral differences as seen in Figure 2, but also may be due to lower-quality data (for instance, those objects with particularly large 2.95-μm band depth). Error bars are 1-σ.

**Figure 9** The 0.7-μm band depth and 2.95-μm band depth for Ch asteroids appear uncorrelated when taken as a whole. Error bars are 1-σ and present in both directions, but they are smaller than the data points for the 0.7-μm band depth..

**Figure 10:** The 0.7-μm band depth and 2.95-μm band depth for the combined set of CM meteorites (with detectable 0.7-μm bands) and Ch asteroids appear to make a single trend. Error bars are omitted for clarity.

**Figure 11:** The 0.7-μm band depth and 2.95-μm band depth for the set of asteroids and meteorites is shown here in two conditions: "Ambient", where the majority of the CM meteorites were measured in a terrestrial atmosphere, and "Dry", which applies an estimated correction to those measurements to simulate their reflectance ratios without terrestrial adsorbed water. The asteroid measurements and meteorite measurements of Takir et al. are unchanged. An apparent trend in the ambient data is largely removed by this correction, leaving a much less significant correlation. Error bars are omitted for clarity.

**Figure 12:** Six Ch asteroids offset from one another, showing a possible absorption near 2.32-2.33 μm of varying depths. The dotted line is a straight-line continuum from the average reflectance at 2.25 to the average reflectance at 2.40 μm. This absorption, if borne out, is consistent with phyllosilicate minerals commonly found in CM meteorites.

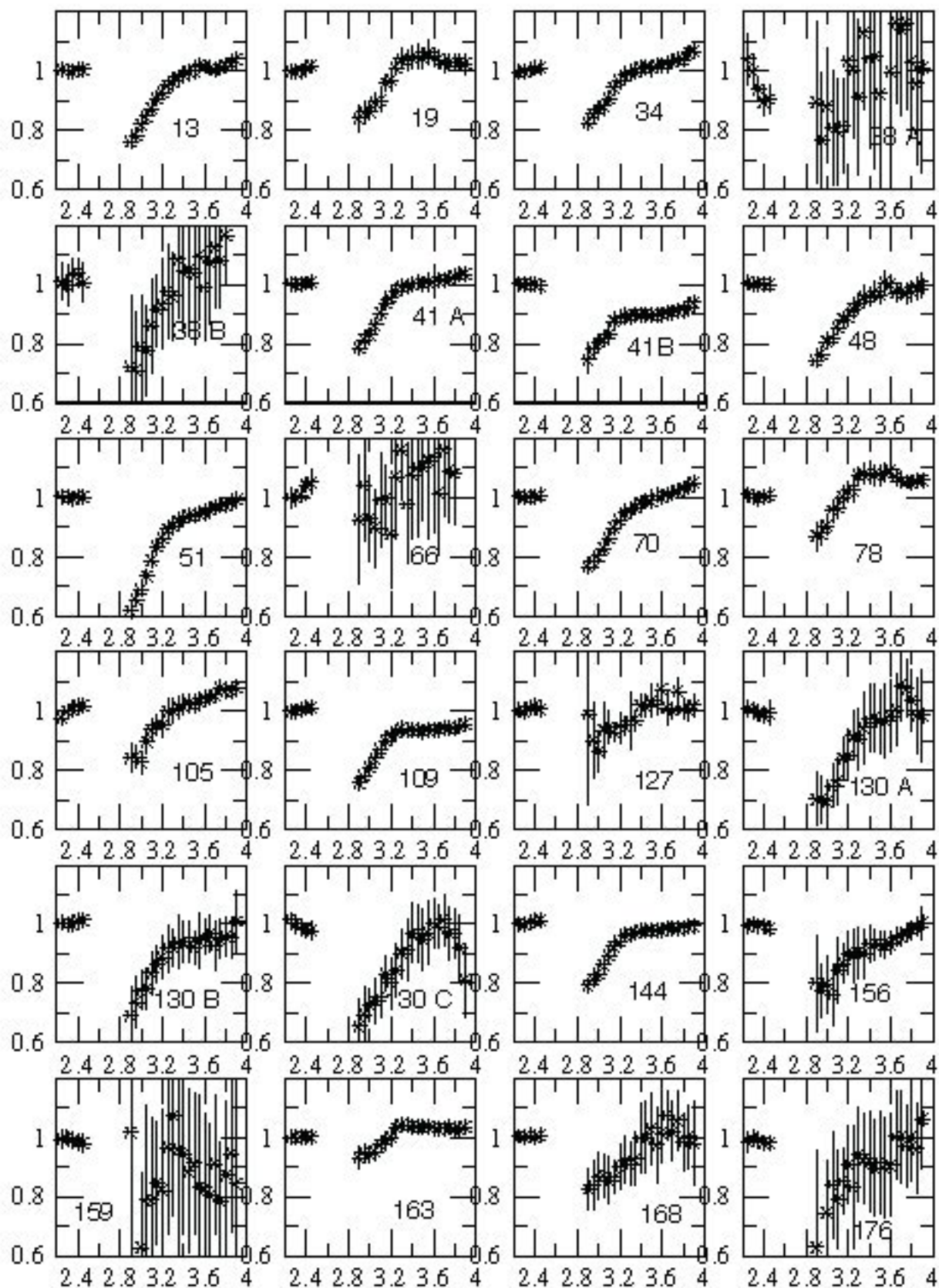

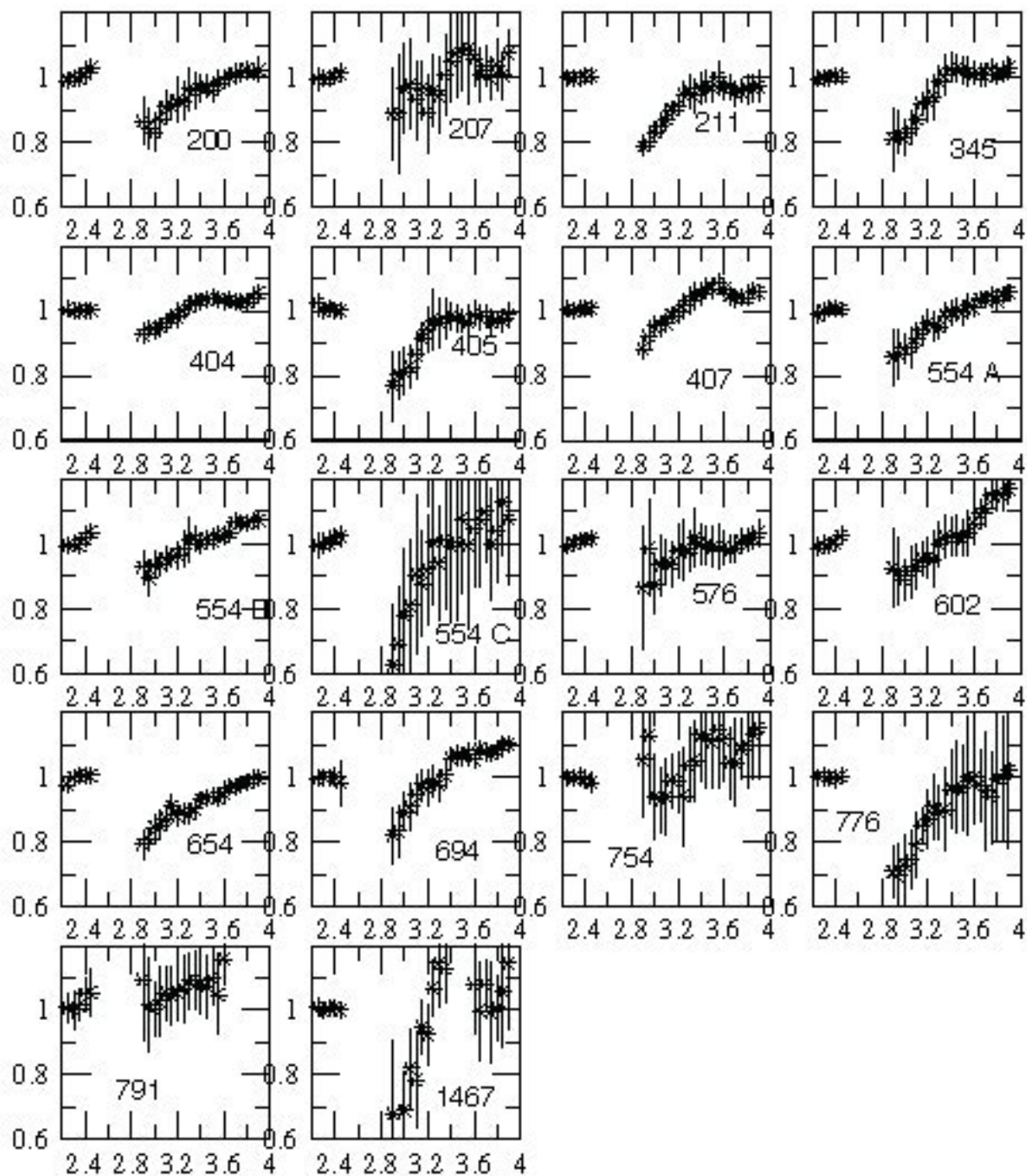

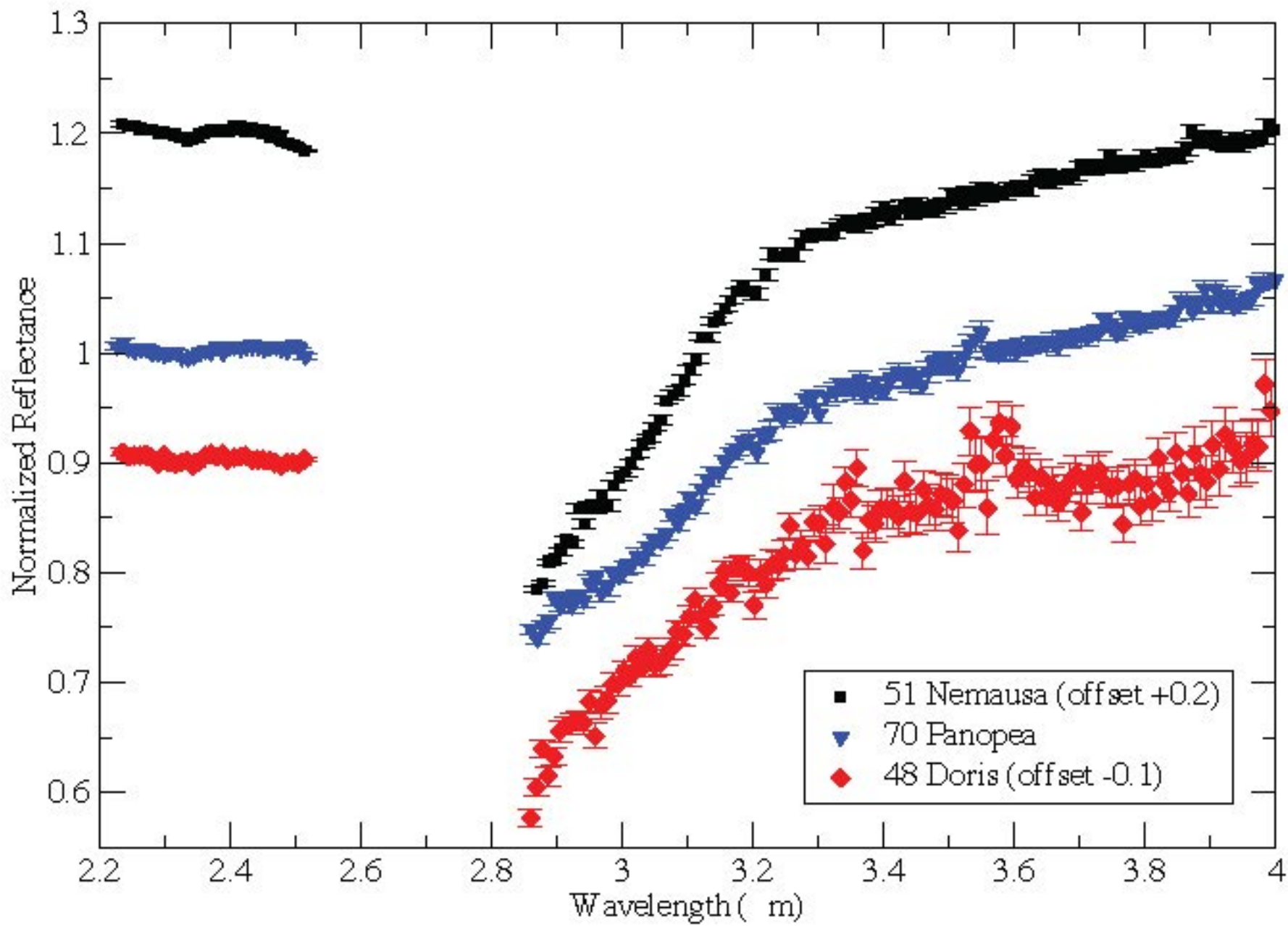

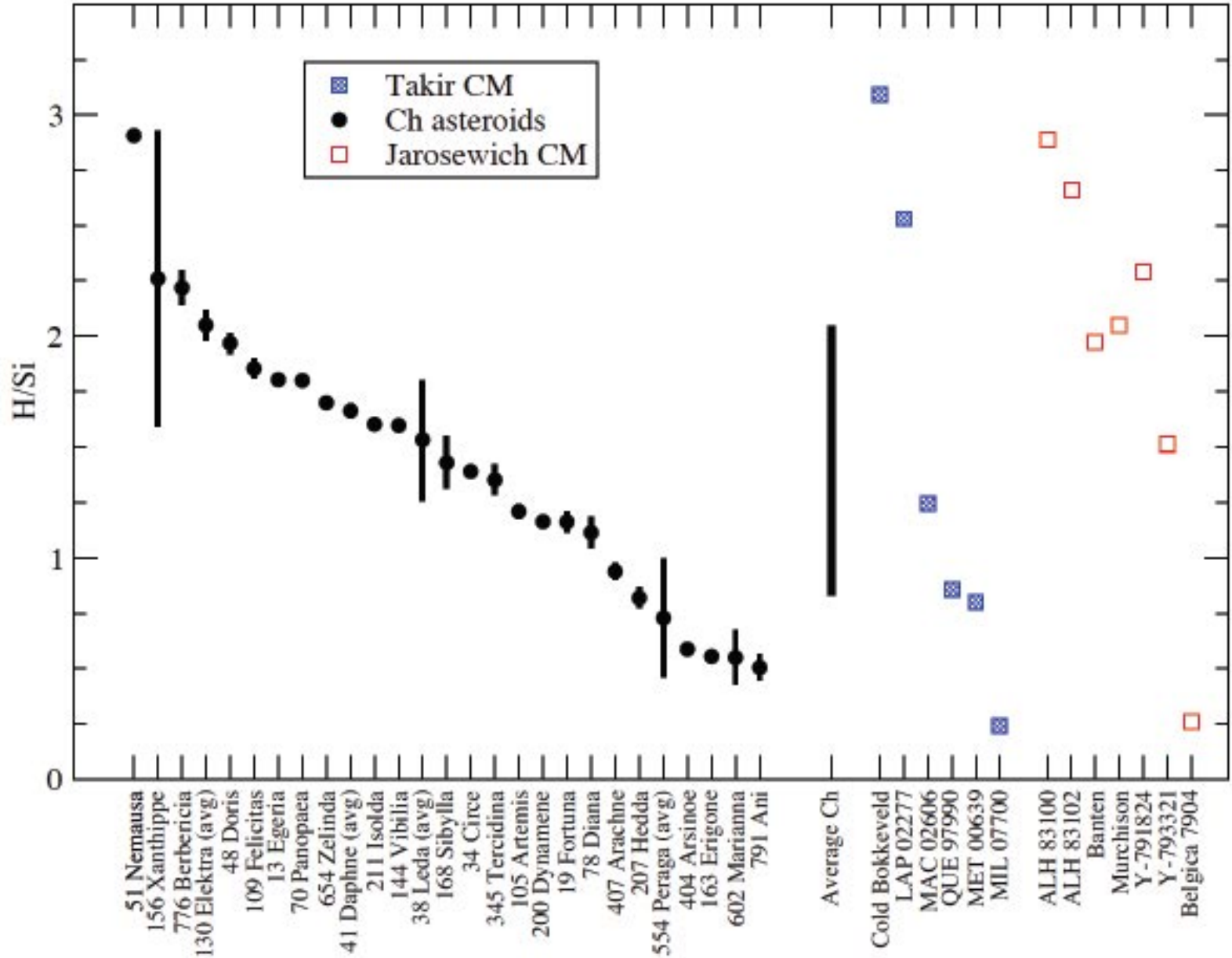

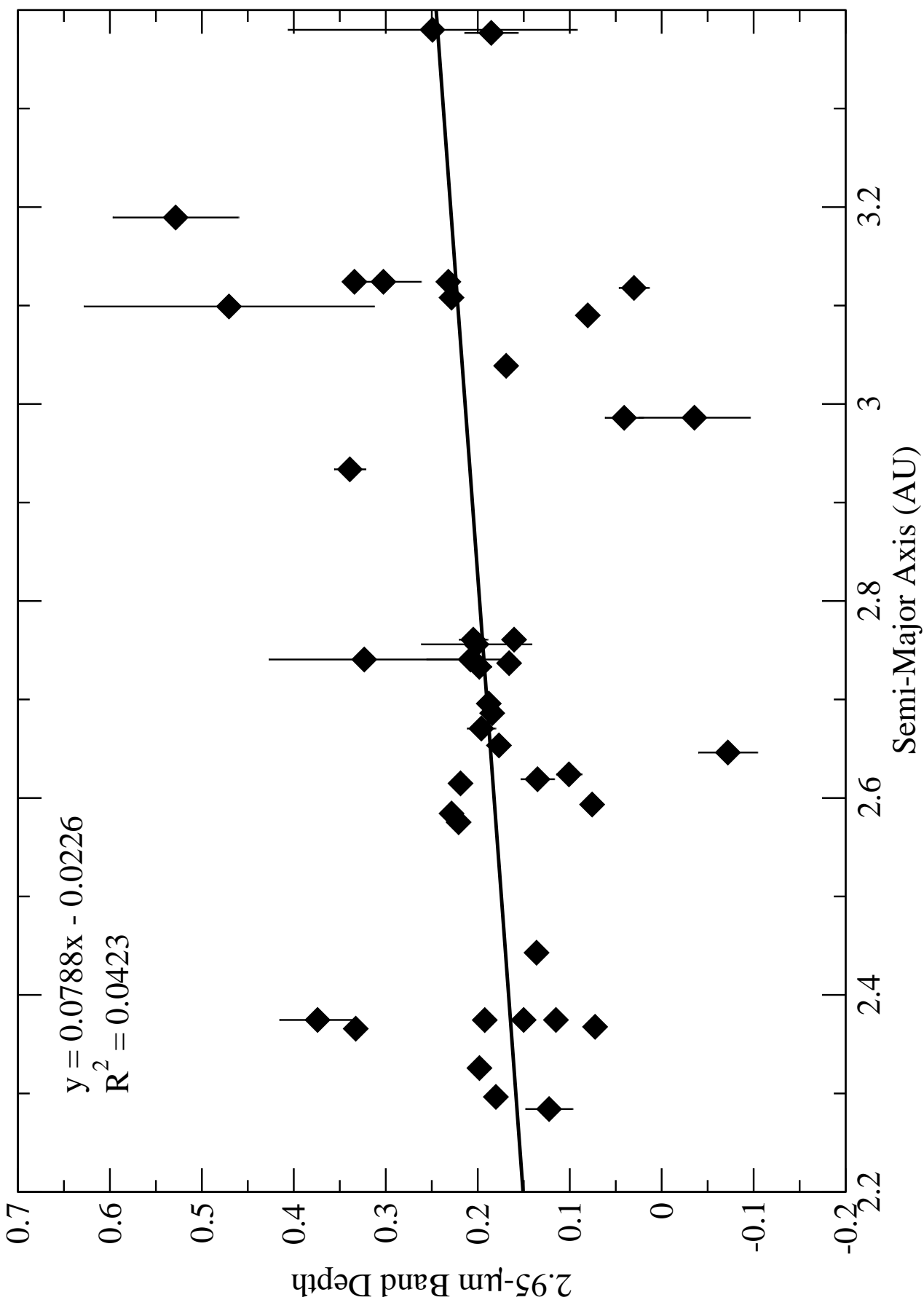

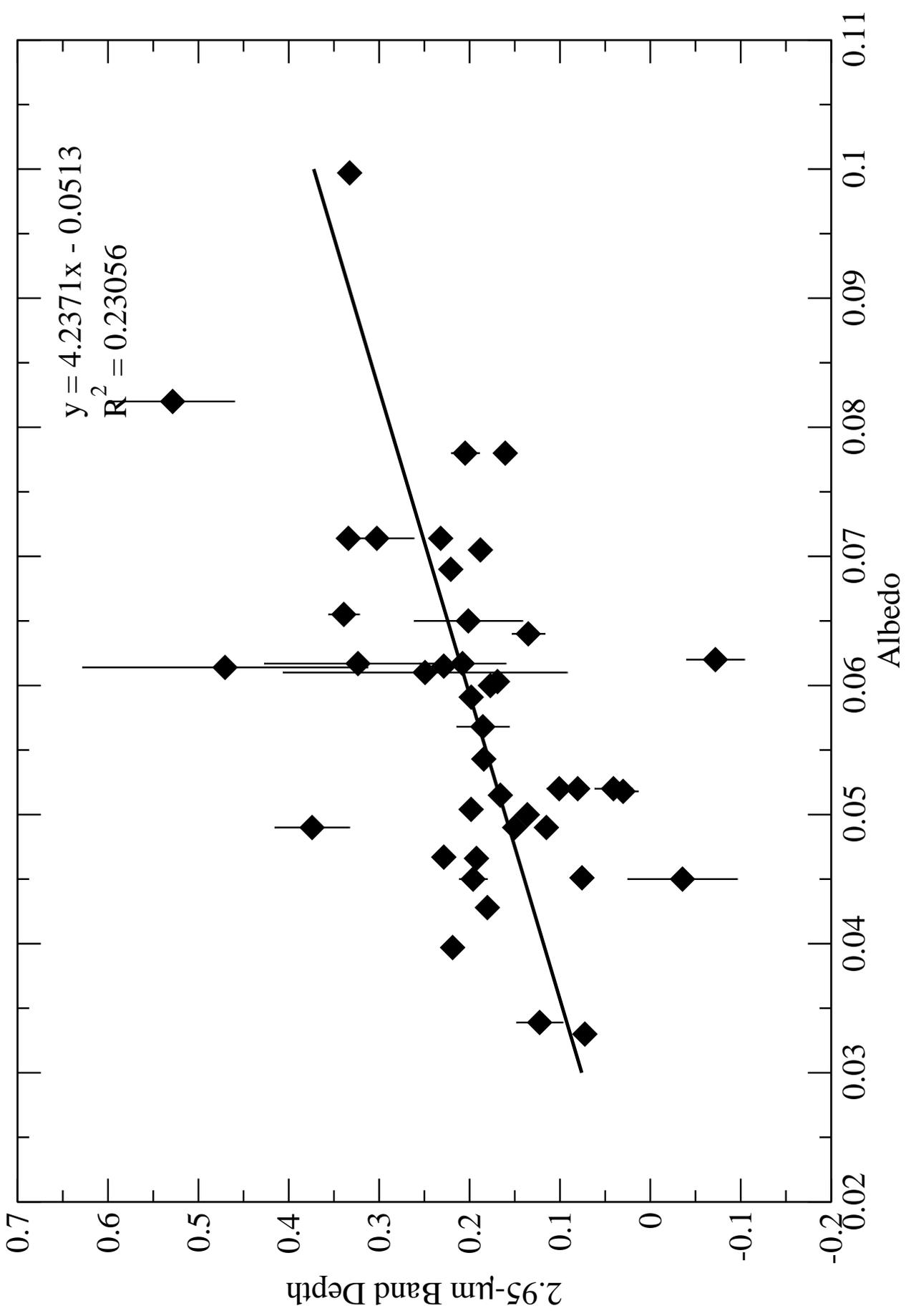

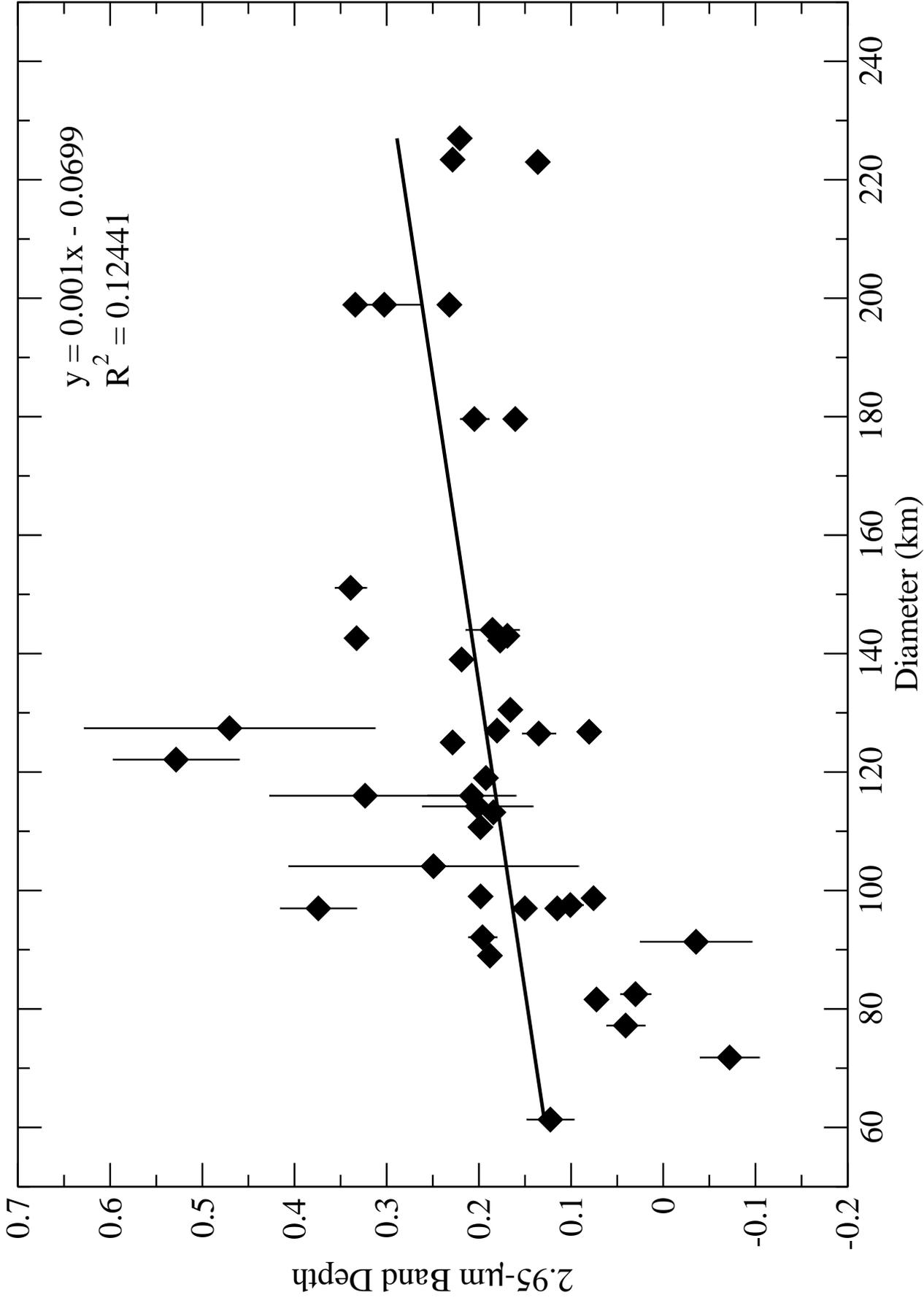

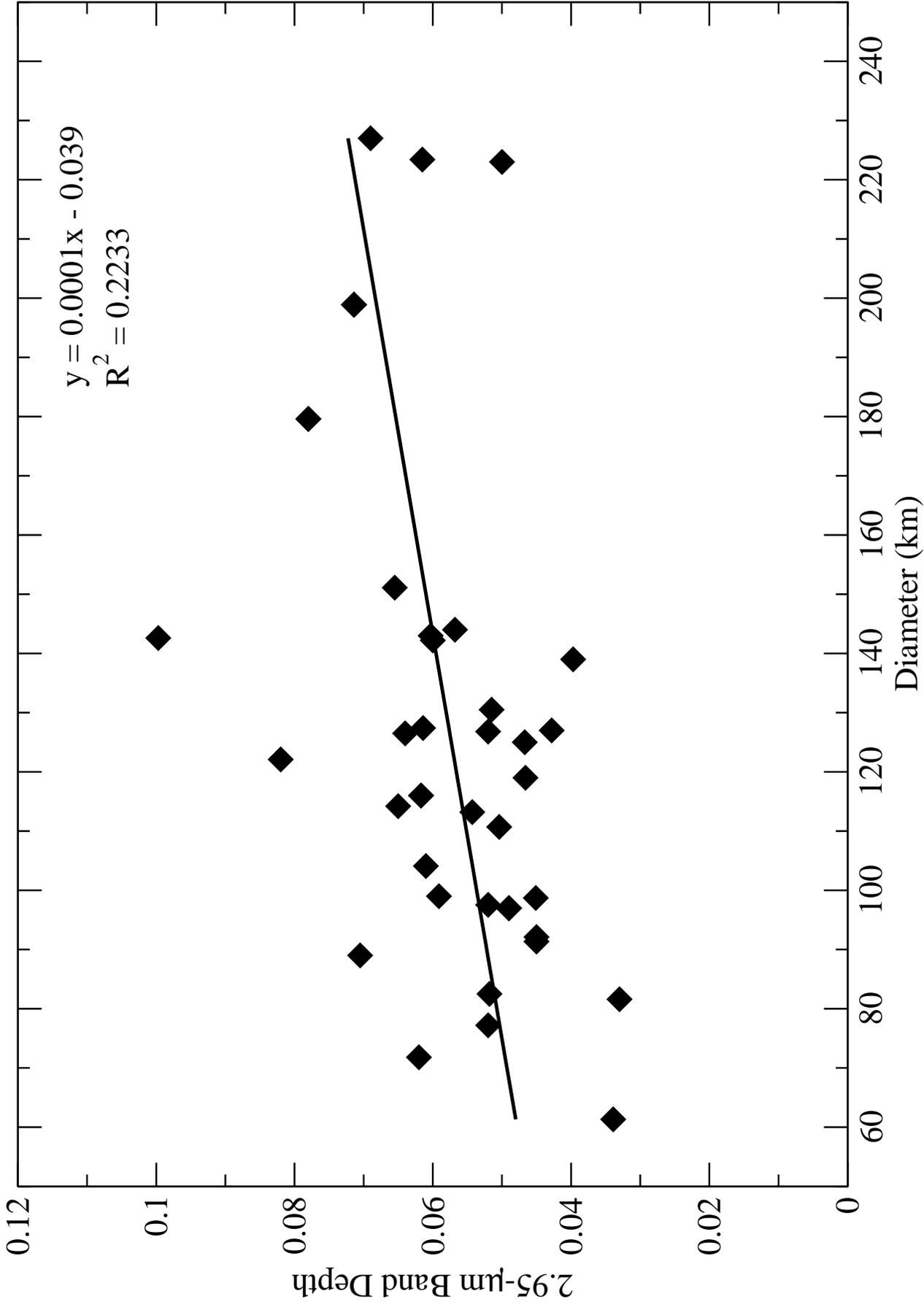

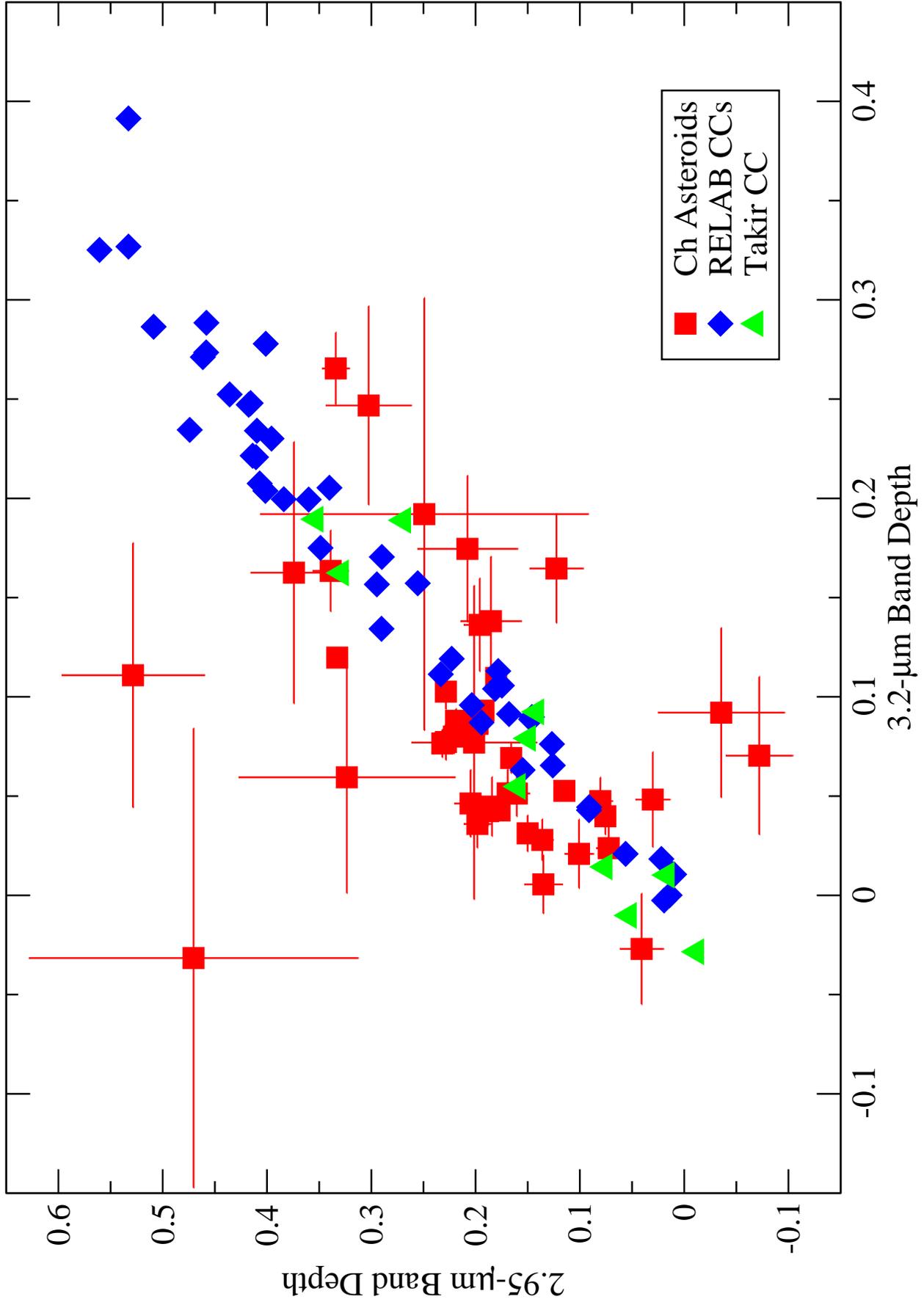

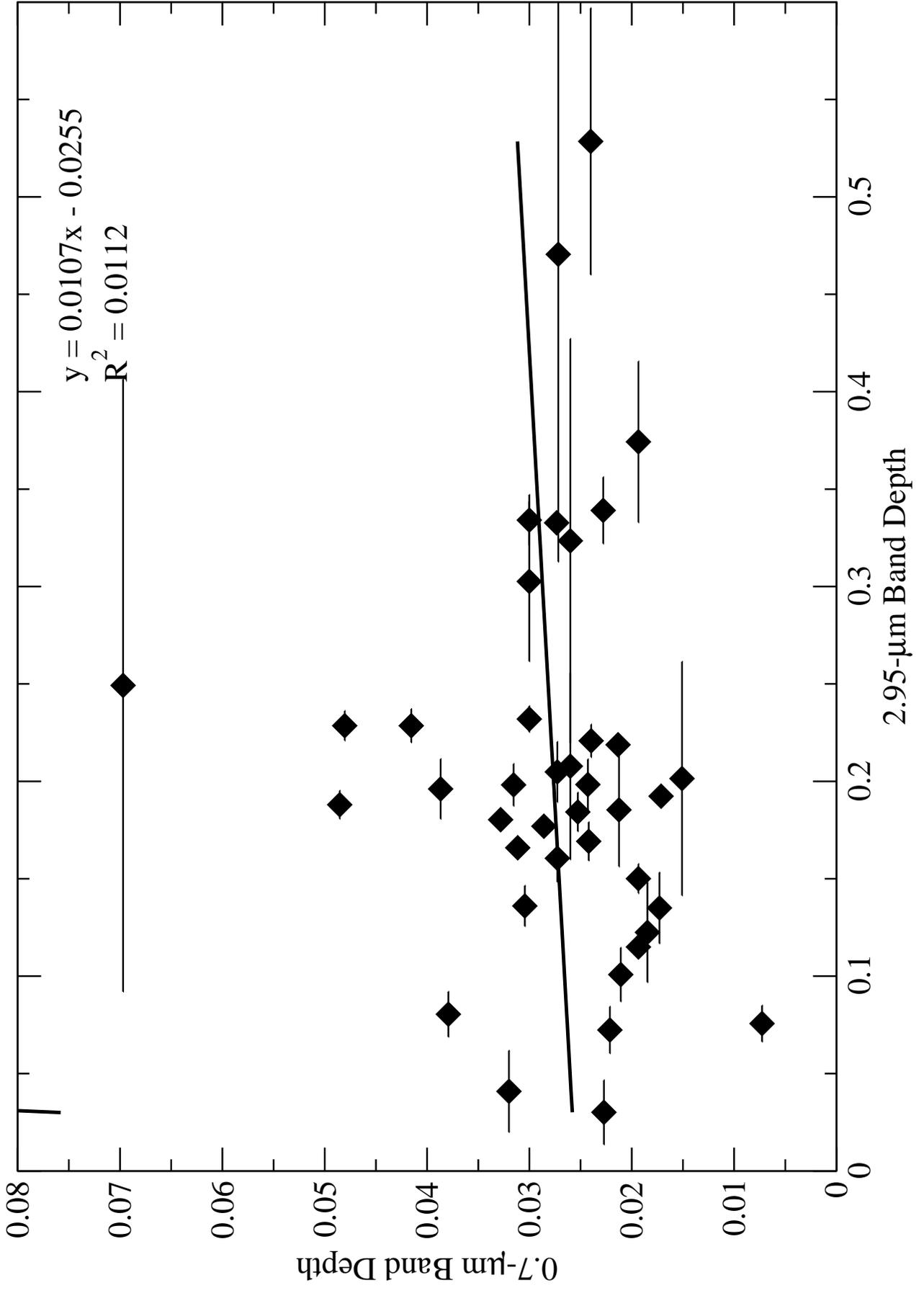

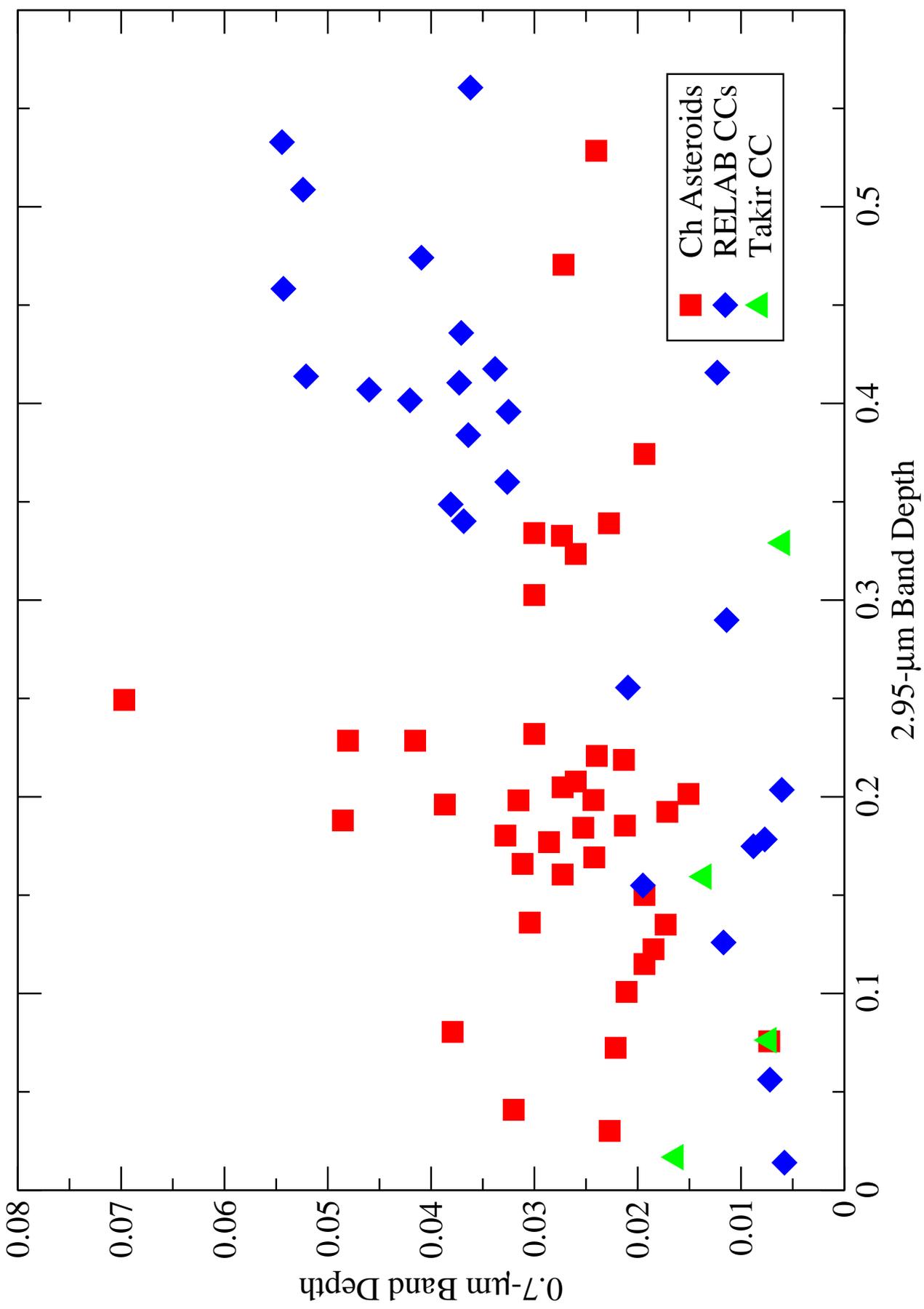

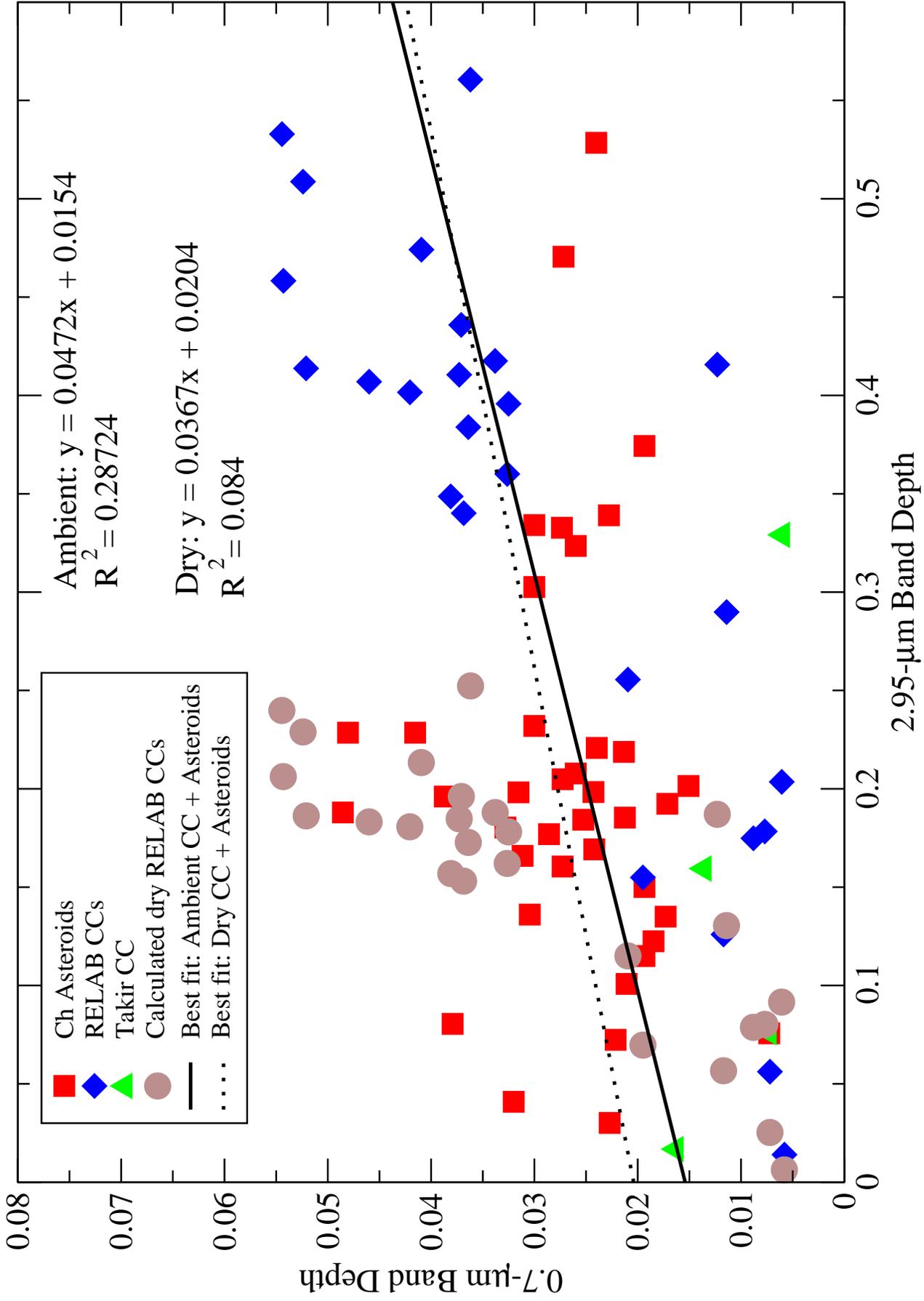

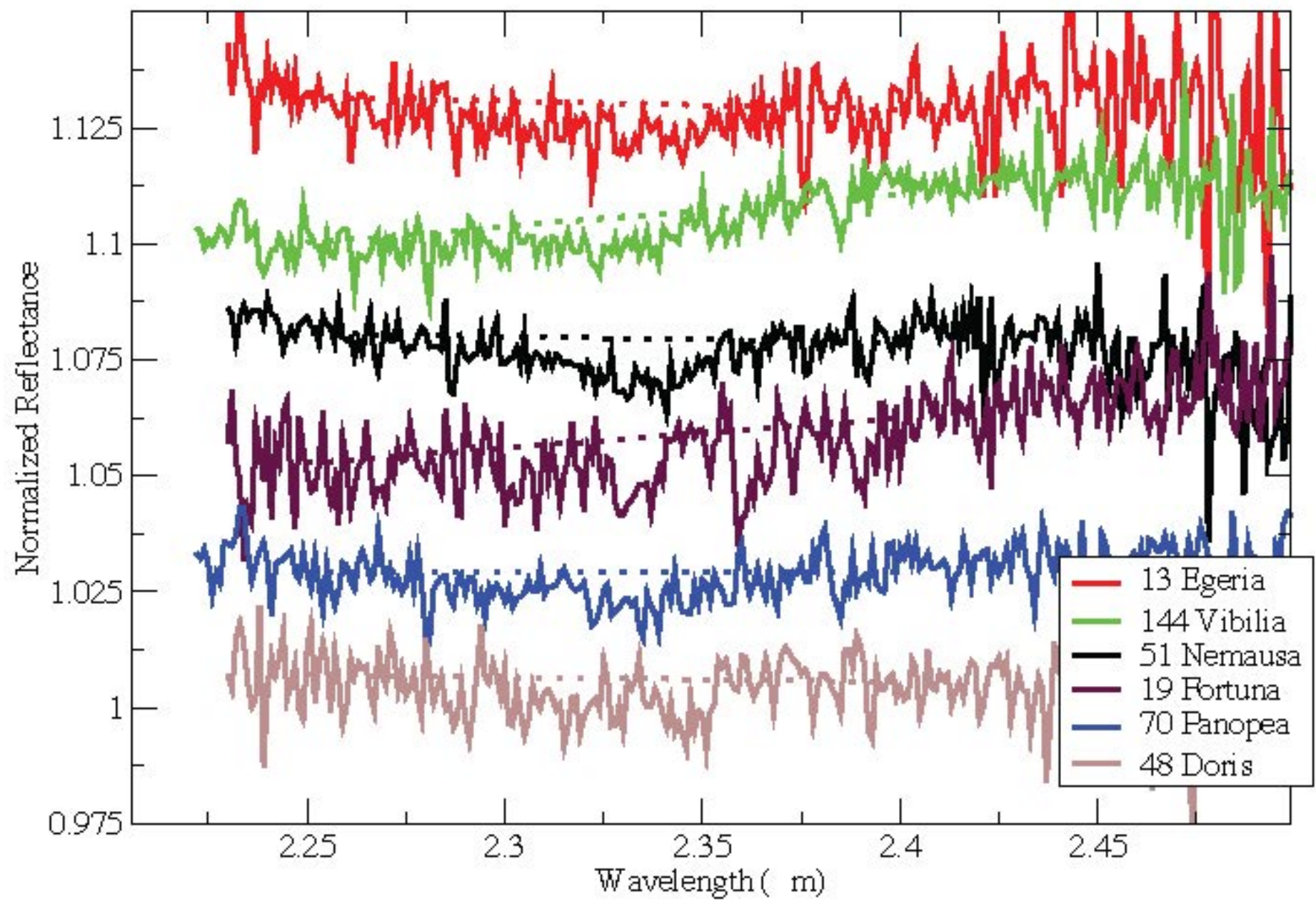